\documentstyle[prl,aps]{revtex}
\input{epsf}

\begin{document}

\title{Front Propagation: Precursors, Cutoffs and Structural Stability}
\author{David A. Kessler and Zvi Ner}
\address{Minerva Center and Dept. of Physics, 
Bar-Ilan Univ., Ramat-Gan, Israel}
\author{Leonard M. Sander}
\address{Dept. of Physics, Univ. of Michigan, Ann Arbor, MI 48109-1120}
\maketitle
\begin{abstract}
We discuss the problem of fronts propagating into metastable and unstable
states.  We examine the time development of the leading edge, discovering
a precursor which in the metastable case propagates out ahead of the front 
at a velocity more than double that of the front and establishes the
characteristic exponential behavior of the steady-state leading edge.
We also study the dependence of the velocity on the imposition of a 
cutoff in the reaction term.  These studies shed new light on the
problem of velocity selection in the case of propagation into an unstable
state. We also examine how discreteness in a particle
simulation acts as an effective cutoff in this case.
\end{abstract}
\section{Introduction}
The study of front propagation is one of the most fundamental problems
in nonequilibrium physics.  As such, it has received much attention
over time, and indeed much is now known about the phenomenon.  For example,
it is known that there is a fundamental difference between propagation into
a meta-stable state (a case we shall label MS in the following) and
that of propagation into an unstable state (US).  
Of the two, the MS case is
the simpler and better understood.  There the front has a unique
velocity which is determined by solving for the unique
traveling-wave solution of the field equations.  The US case is more
subtle.  There the same procedure produces a continuum of possible
velocities and associated steady-state solutions.  Thus the velocity
selection is not a result simply of the steady-state equation as
it is in the MS case.  The question of how
a particular velocity is selected is then paramount.  Much progress has
been made on this question since the pioneering work of Kolmogorov, et. al.
\cite{Kolmo}
in the 1930's and Aronson and Weinberger 
\cite{AronWein} some
twenty years ago.  Nevertheless, a good intuitive physical picture has
been lacking.  Noteworthy attempts to supply such a picture include
the famed "marginal stability" hypothesis of Langer and collaborators
\cite{Dee,Ben-Jacob} and the "structural stability" hypothesis of
Goldenfeld and collaborators \cite{Gold1,Gold2,Gold3}.  

Most recently, the
work of Brunet and Derrida (BD) \cite{BD} has cast a most illuminating
light on this problem.  They focus on two questions.  The first is
the time development of the front.  They solve the problem of how the
steady-state front is developed starting from a typical initial condition.
They show that the front slowly builds itself up over time, so that the
tail of the steady-state front extends farther and farther ahead of the
bulk of the front as time goes on.  This slow build up leads to a very
slow approach of the front velocity to its steady-state value, with a 
correction that vanishes only as $1/t$ as the time $t$ goes to infinity.
The second question BD address is the effect on the velocity of
the front of a cutoff $\epsilon$
on the reaction
term far ahead of the front\cite{RNA}.  
They show that this effect is surprisingly strong. First
of all, any cutoff, no matter how small, serves to select a unique
velocity, which approaches a finite limit as the cutoff is removed.  
This is indicative of a structural instability of the original problem,
and indeed is the basis for the considerations of the structural
stability hypothesis of \cite{Gold1,Gold2,Gold3}.
Moreover, the dependence of this selected velocity on the cutoff is
quite dramatic, approaching its limiting value only as
$(\ln \epsilon)^{-2}$.

The purpose of this paper is to further develop the insights into
front propagation afforded by the work of BD.  The two issues raised
by BD are of importance for all front propagation problems, and not
just for the US problem.  In particular, looking at the buildup
of the front in the MS case reveals the hitherto unnoticed existence
of a ``precursor'' front.  The role of this ``precursor'', which moves more
than twice as fast as the ``bulk'' front itself, serves to build up
the characteristic exponential leading edge of the steady-state front.
Turning back to US, we then see that in this case such a precursor is 
impossible, except for the special selected velocity.  We thus see
that all but this special velocity are not realizable from localized
initial conditions.  In addition, we can understand
BD's solution as a special degenerate case of the general precursor.

Similarly, examining the effect of a cutoff in MS, we see that the
induced velocity shift is weak, and in fact superlinear.  
This is in contradistinction
to the strong $(\ln \epsilon)^{-2}$ behavior seen by BD in the US case.
Again, this difference can be tied to the different structure of the
linearized problems in the leading edge.  The strong $\epsilon$ dependence
in the US case is seen to be a reflection of the structural instability
of this problem.

In fact, we discover that both of BD's two issues are intimately related.  
The effect of the
localized initial conditions acts as an effective cutoff for the
steady-state dynamics.  This cutoff is time-dependent, and vanishes
as more and more of the steady-state leading edge is built up.  The
difference between the familiar exponential relaxation of the velocity in MS
and the $1/t$ relaxation in US is, as we shall see, a result of the very 
different sensitivities to the cutoff in the two cases.  Thus, we may
alternatively understand the ``weak'' velocity selection of the US
case as being driven by the initial conditions, as opposed to the ``strong''
velocity selection of MS, which results directly from the steady-state
equation.

In addition, we examine in more detail the
conjecture of BD that in a discrete system of reacting particles, the 
discreteness acts as an effective cutoff.  We find that, while qualitatively
correct, a deeper understanding is needed to make this association
quantitatively reliable. Lastly, we study the US case called
"nonlinear" or "Type II" marginal stability and show that it
interpolates between the typical US case and MS.  We also find that
an examination of the linear stability of the steady-state solutions
in this case sheds a new
light on the meaning of the ``marginal stability'' hypothesis.

\section{Propagation into a Meta-stable State: The Ginzburg-Landau Equation}
The most classic example of front propagation into a meta-stable
state is the Ginzburg-Landau
equation:
\begin{equation}
\label{gl}
\dot\phi = D \phi'' + (1-\phi^2)(\phi + a)
\end{equation}
This equation possesses two locally stable states at $\phi=\pm 1$ with an
unstable state at $\phi= -a$.  When started with an initial condition
such that the system is in the $\phi = -1$ state except for some 
region at the left which is in the $\phi=1$ state, the $\phi = 1$ state
will propagate into the $\phi=-1$ state (for $0<a<1$).
The propagating front quickly achieves 
a time-independent shape which moves at a constant velocity
$c=c(a)$.  This velocity and front shape are given by the unique solution of
the steady-state equation
\begin{equation}
\label{glss}
D \phi'' + c \phi' + (1-\phi^2)(\phi + a) = 0
\end{equation}
where now $\phi$ is a function of $y\equiv x-ct$ and satisfies the
boundary conditions $\phi(y\to \mp\infty)=\pm 1$. This velocity selection
can be simply understood from a classic mode counting argument.  As
$y\to -\infty$, Eq. (\ref{glss}) for a given $c$, linearized around $\phi=1$,
 possesses
two exponential modes, one decaying and one growing.  This is a direct
result of the stability of the $\phi=1$ state. Thus, there is
one degree of freedom in the solution at $y=-\infty$, the fixing of which
corresponds to breaking the degeneracy induced by translation symmetry.
Doing this leaves no freedom in the solution at $y=-\infty$.  Now, as
$y \to \infty$, since the $\phi=-1$ state is also stable, there are again 
two exponential modes, one decaying and
one growing.  The requirement that the growing mode be absent from the
solution fixes the velocity. This behavior of the linearized equation
will be the crucial ingredient in all that follows.

For the cubic reaction term in Eq. (\ref{gl}), the explicit form of the
solution can be obtained, and is given by
\begin{equation}
\phi=-\tanh(q(y-y_0)/2)
\end{equation}
with $c=a\sqrt{2D}$, $q=(c+\sqrt{c^2 + 8D(1-a)})/2D$ and $y_0$ arbitrary.  
However, as we have indicated, what is important for us is
the asymptotic behavior of the solution
for large $y$, which is $\phi \sim -1 + A\exp(-qy)$ with the
constant $A$ dependent on $y_0$.  

\begin{figure}
\centerline{\epsfxsize=3.25in \epsffile{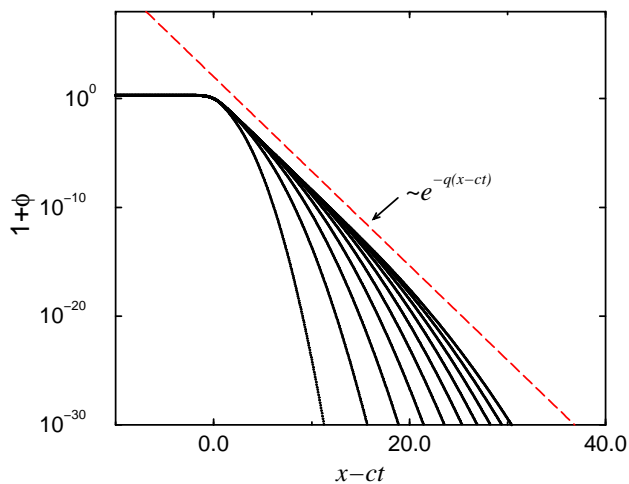}}
\caption{$\phi+1$ vs. $y=x-ct$ for GL eqn. (Eq. 1), $a=0.1$, $D=0.5$ 
where $\phi(x,t=0)=1-2\theta(x)$, for $t=1,\ 2,\ \ldots ,\ 10$.}
\end{figure}

The first of BD's
two questions we
wish to address is how this front develops itself.  To begin,
let us examine the result of a simulation.  We started with a step-function
front, and integrated forward in time.  In Fig. 1, we plot $1 + \phi$ in
the frame moving with the asymptotic front velocity $c(a)$, for different
times.  We see that in the front region
$1+\phi$ approaches the pure exponential behavior noted above. 
However, far ahead of the moving front, $1+\phi$
falls much faster than exponentially.  The transition region where the
front starts to deviate from the pure exponential is seen to move steadily
ahead, this in the co-moving frame, leaving in its wake more and more
of the correct exponential front.  There is thus seen to be a ``precursor''
front, which acts to build up the true steady-state front.

Motivated by the numerical evidence and the results of BD, we 
assume the following ansatz for $1+\phi$ in the leading edge, in
terms of the moving frame variable $y$:
\begin{equation}
\label{precursor}
1+\phi \sim A e^{-qy} f(\frac{y-c^*t}{t^\alpha})
\end{equation}
Since we are interested in the front region, where $1+\phi$ is small,
we can linearize Eq. (\ref{gl}) about $\phi=-1$.  Of course, the linear
equation does not know which $q$, (or equivalently $c$) to choose, since
this is fixed by the full nonlinear problem.  Nevertheless, we know we
must choose the steady-state $c$, $q$ in Eq. (\ref{precursor}) if we are to 
match on to the bulk steady-state solution.  Substituting Eq. 
(\ref{precursor}) into the 
linearized version of Eq. (\ref{gl}), and
writing $z\equiv (y-c^*t)/t^\alpha$ we find 
\begin{equation}
f'(-c^*/t^{\alpha} - \alpha z/t)=D(f''/t^{2\alpha} - 2qf'/t^{\alpha}) + cf'/t^{\alpha}
\end{equation}
We see that we can eliminate the leading order $t^{-\alpha}$ terms, if we
choose
\begin{equation}
c^* = 2qD - c = \sqrt{c^2 + 8D(1-a)}
\end{equation}
Then, the other terms balance if we choose $\alpha=1/2$, resulting in the
following equation for $f(z)$:
\begin{equation}
D f'' = - z f'/2
\end{equation}
whose solution such that $f(z \to -\infty)=1$ is 
\begin{equation}
f(z) = \frac{1}{2} \text{erfc}(z/2\sqrt{D})
\end{equation}
We show in Fig. 2 a plot of $\exp(qy)(1+\phi(y))$ together with our 
analytical result.  The agreement is clear.

\begin{figure}
\centerline{\epsfxsize=3.25in \epsffile{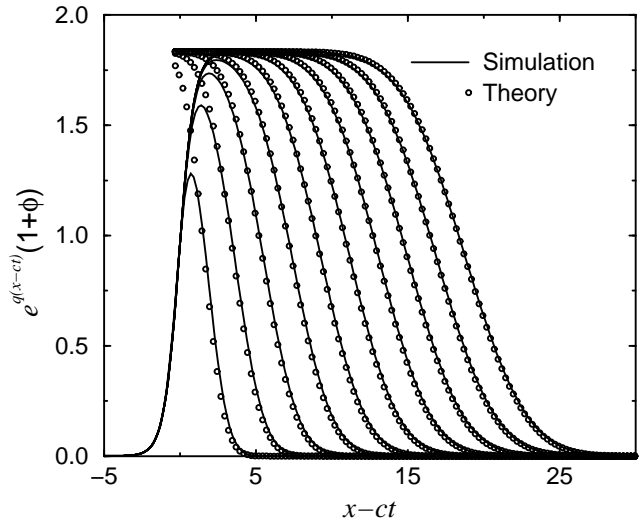}}
\caption{$\exp(qy)(\phi +1)$ vs. $y=x-ct$ for same simulation as in Fig. 1.
Theory curve is $A\,\text{erfc}((y-c^*t)/2\protect{\sqrt{Dt}})$, with 
$A=1.8535$
a fitting parameter, and the translation invariance is used to suitably adjust
the origin of both $y$ and $t$.}
\end{figure}

Let us make a few observations regarding this result.  First we note
that, depending as it does only on the linearized equation, it is extremely
general.  Second, we see that the precursor front indeed moves ahead of the
front, with a constant velocity in the lab frame of $c + c^*> 2c$.  
The precursor
front exhibits diffusional broadening, so that its width increases as 
$t^{1/2}$. For fixed $y$, large $t$, we find that
\begin{equation}
\label{larget}
1+\phi \sim A e^{-qy} - A \sqrt{D} e^{-cy/2D}\, \frac{e^{-(c^*)^2t/4D}}{c^*\sqrt{\pi t}}
\end{equation}
This is directly related to the structure of the stability operator for
the steady-state solution.  The spectrum consists of a discrete zero
mode and a continuum of states with decay rates that extend from
$(c^*)^2/4D$. The $t^{-1/2}$ results from the integration over the continuum
of states, which has the structure $\int dk \exp(-k^2 t)$. The factor
$\exp(-c y /2D)$ is just the decay rate of these continuum states. We can also
read off from this result the approach of the velocity to its asymptotic
value:
\begin{equation}
c(t) - c \propto \frac{e^{-(c^*)^2t/4D}}{\sqrt{t}}
\end{equation}
The prefactor depends on the exact definition of the front
position, and in fact also varies in sign.  If the front position is defined
by the point where $\phi$ crosses some particular value, than if this value
is chosen sufficiently close to $-1$, (in particular, if it is close enough
that the linear solution above is relevant to describe the front position)
the coefficient is always positive 
so that the velocity approaches its asymptotic value from above. 

The second of BD's two questions is the response of the system to
the imposition of a cutoff in the reaction term\cite{RNA}.  We modify the
reaction term to $(1-\phi^2)(\phi + a)\theta(1+\phi-\epsilon)$ such that
the reaction term is turned off if $\phi$ is too close to its meta-stable
value of $-1$.  In the presence of the cutoff, there is still a unique
selected velocity, which is now $\epsilon$-dependent, which we write
$c_\epsilon = c + \delta c$. To see this, it is sufficient to note that, as
in the $\epsilon=0$ case, up to the translation zero-mode, there is no 
freedom of the steady-state solution (with some given velocity) as 
$y \to -\infty$. We integrate
the solution forward in $y$ till the point $y_c$ when it reaches 
$\phi=-1+\epsilon$.  The
first derivative of $\phi$ at this point must match
on to the unique solution of the steady-state equation in the
non-reacting region, $y>y_c$:
\begin{equation}
\label{noreact}
\phi=-1 + \epsilon \exp(-c_\epsilon (y-y_c) / D) \ .
\end{equation}
This requirement fixes the velocity. 

We may calculate $\delta c$ for small $\epsilon$ as follows.  For large
$y < y_c$ we may linearize the equation about $\phi=-1$.  The general
solution is, to leading order,
\begin{equation}
\label{mstail}
1 + \phi = A e^{- q y} + B \delta c e^{l y}
\end{equation}
where $q$ is as above and $l=(-c+\sqrt{c^2 + 8D(1-a)})/2D$ is the spatial 
growth rate of the second, growing mode of the linearized equation. 
The coefficients $A$ and $B$ are fixed by
matching this to the solution of the steady-state equation, treating 
$\delta c$ as a perturbation.  What is important is that 
$A$ and $B$ are independent of $\epsilon$.  We have to match this
solution at $y_c$ to the non-reacting solution Eq. (\ref{noreact}) above.
This implies that $q y_c$ is of order $\ln(1/\epsilon)$, and that therefore
\begin{equation}
\delta c \sim \epsilon^{1 + l/q}
\end{equation}
Note that $0 < l/q < 1$, so that $\delta c$ vanishes faster than linearly,
with a power between 1 and 2.  This result is corroborated by
a numerical solution of the cutoff steady-state equation, presented in
Fig. 3.  The weakness of the effect of
the cutoff on the velocity is a result of the {\em structural stability} 
\cite{Gold1,Gold2,Gold3} of the velocity selection problem in this system.  
Both with and without $\epsilon$, the nature of the velocity 
selection problem is the same:
only a precise dialing of the velocity will achieve the desired behavior
in the front region.  This is because in both cases the dying exponential 
behavior of the front is {\em unstable}, an infinitesimal change of the
velocity will induce a different dominant behavior of the front. Furthermore,
since the dying exponential behavior of the $\epsilon=0$ front is unstable, 
we can make major, i.e. 
order 1, changes in it via relatively minor changes in $c$.

Before leaving the GL equation, we wish to make one additional
remark.  We noted above the connection between the structure of the
precursor front and the stability analysis.  In particular, the GL
stability operator had no discrete modes other than the translation
zero mode.  However, it is easy to modify the reaction term so that
the stability operator has some finite number of discrete stable modes
without dramatically changing the overall structure of the selection
problem.  Clearly, the long-time behavior of the front region is dominated
by this mode, and it is important to understand how the front region
develops in this case.  In the case where the stability operator has a single
stable discrete mode with decay rate $\Omega$, for example, the 
generalization of Eq. (\ref{precursor})
is given by
\begin{equation}
\label{2precur}
\phi - \phi(y=\infty) \sim A e^{-qy} f(\frac{y-c^*t}{\sqrt{t}})
+ B e^{-\Omega t} e^{-q^\Omega y} f(\frac{y-c^{\Omega}t}{\sqrt{t}})
\end{equation}
where $q^\Omega$ is the spatial decay rate of the discrete mode, and $c^\Omega=
2q^\Omega D - c$, and $f$ is as above.  This can be shown to also
satisfy the linearized dynamical equation for $\phi$, and goes over
at long times and fixed $y$ to a sum of the steady-state (front) solution
$A \exp(-qy)$ plus the decaying stable mode.  Again, the linear equation
does not fix $\Omega$, which we choose so as to match onto the
solution of the full stability
operator.  We see from Eq. (\ref{2precur}) that in this case there
are 2 precursors, one associated with the solution and one with the stable
discrete mode.  As $q^\Omega < q$, the velocity $c^\Omega$ of the latter
is always slower than the velocity $c^*$ of the former.  
In fact, one can show that the stable mode precursor is always sufficiently
slow that the stable mode is cut off by $f$ before its slower spatial
decay rate allows it to overcome the steady-state front.
Thus the finite propagation speed of the stable-mode is crucial in
ensuring that the entire front is dominated by the steady-state and not
by the perturbation.

\begin{figure}
\centerline{\epsfxsize=3.25in \epsffile{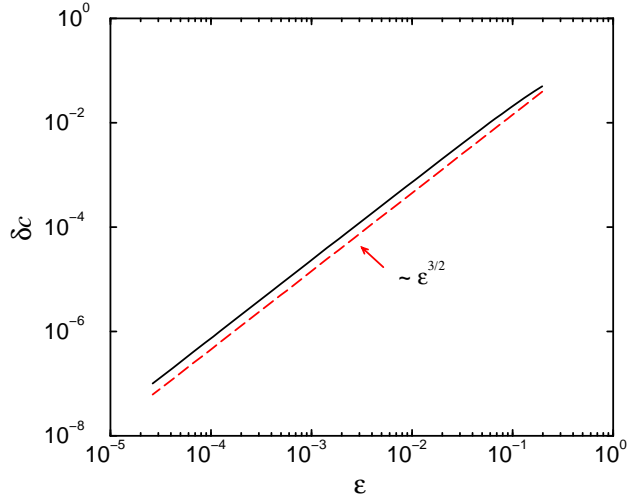}}
\caption{$\delta c\equiv c(\epsilon) - c(\epsilon=0)$ vs. $\epsilon$ 
for MS propagation in the GL equation, Eq. (\ref{gl})
with $D=1/2$, $a=1/2$.  The
theoretical scaling prediction $\delta c \propto \epsilon^\beta$,
$\beta = 1+ l/q=3/2$ is also shown.}
\end{figure}

\section{Propagation into an Unstable State: Fisher equation}
We can now use the results of the above section on the behavior of the 
GL equation to give a broader context to, and thus better understand,
the behavior of the Fisher equation\cite{Fisher}.  The Fisher equation 
(also referred to in some circles as the Fisher-Kolmogorov, or 
Fisher-Kolmogorov-Petrovsky-Piscounov equation) is
\begin{equation}
\dot \phi = D \phi'' + \phi(1-\phi)
\end{equation}
Again, we have a stable state at $\phi=1$, but now there is no
second stable state, just an unstable state at $\phi=0$.  If we start out
with a step function initial condition interpolating between these
two states, then the stable state propagates into the unstable state
at a velocity $c=2\sqrt{D}$.  The problem is that the steady-state
equation possesses solutions for all velocities.  As is the MS
case, this follows from simple mode counting.  As in MS, since the
solution approaches the stable state at large negative $y$, there is
one growing and one dying exponential mode of the linearized operator.
Again, fixing the translation mode leaves us with no freedom of the
solution as $y \to -\infty$.  Now, however, in contradistinction to MS,
the instability of the $\phi=0$ state at large positive $y$ implies
that both modes of the linear operator are dying exponentials. There
is then no growing mode to eliminate, and all velocities are acceptable.
Nevertheless, a step-function initial condition leads to a selected
velocity of $2\sqrt{D}$. Understanding
this ``dynamical'' selection of the chosen velocity is a challenge, one
that has been taken up by many different groups over the years.

Let us first discuss the time-dependence of the front.  If we
attempt to apply the technique we used in the GL case, we run into an
immediate problem.  The velocity of the precursor front, in the co-moving
frame, is again $c^*=2Dq - c$.  However, now $q$, the spatial decay rate of the
front, is given by the slower of the two dying exponentials, and
so is $q=(c - \sqrt{c^2 - 4D})/2D$.  Thus, $c^*=-\sqrt{c^2 - 4D}$
is {\em negative}, which clearly makes no sense.  Thus, there is in
general no way
to build up the appropriate exponential front.  The only
way out is if $c=2\sqrt{D}$, so that the precursor moves at the same velocity
as the front itself.  This then is the origin of the ``dynamical''
selection; fronts with faster velocities, and hence slower spatial decay
cannot build themselves up
{\em ex nihilo}, although they are perfectly stable (in an infinite
system; a finite system will provide an effective cutoff, see below)
if they are 
present in the initial condition.  Slower fronts are linearly
unstable\cite{Ben-Jacob}, and are not dynamically relevant, 
as can be seen directly from
Eq. (\ref{larget}), since $c^*$ is imaginary in this case and so the front 
grows exponentially in time.

The dynamics of the developing front is thus seen to be degenerate
and requires a slightly more sophisticated analysis, which is exactly 
that provided
by BD. They find that the front development is given by
\begin{equation}
\phi=A z e^{-(z^2)/4Dt}e^{-z/\sqrt{D}}
\end{equation}
where $z=y - 3/2 \ln t$ is the front position.
Again, this function describes a spreading Gaussian far ahead of 
the front.  For fixed $z$, as $t$ becomes large, the true steady
front is reproduced.  Here, however, since $c^*=c$, the steady
front is only achieved through diffusive spreading and thus only 
for a distance of order $t^{1/2}$ ahead of the
front.  This is a reflection of the marginal stability of the 
$c=2\sqrt{D}$ front.

We now turn to the effect of a cutoff on the Fisher equation, now
turning off the reaction term when $\phi < \epsilon$.  Here,
the cutoff changes {\em qualitatively} the nature of the steady-state selection
problem.  Whereas for $\epsilon=0$ the steady-state equation possessed
solutions for {\em all} velocities, any finite $\epsilon$ implies that
only a single velocity is possible.  This is a result of the fact that,
whereas for $\epsilon=0$, the decaying nature of the solution is stable,
with finite $\epsilon$, it is not, as we saw in the GL equation.  This
was first pointed out by Goldenfeld, et. al., \cite{Gold2} 
who also noted that $c_\epsilon$
approaches $2\sqrt{D}$ as $\epsilon$ is taken to 0.  They also presented
data to this effect, but did not investigate the nature of the approach,
which was studied by BD. BD found that the
deviation from the asymptotic value $2\sqrt{D}$ is anomalously large, vanishing
only as $(\ln \epsilon)^{-2}$.

Here we wish to emphasize the difference between the Fisher case
and that of the GL discussed above.  In the GL case, the instability
of the decaying behavior implied that a small change in velocity allowed
for an $O(1)$ change in $\phi'/\phi$.  In the Fisher case, for velocities
larger than $2\sqrt{D}$, the decaying behavior is stable, and so small
changes in velocity modify $\phi'/\phi$ by a small amount from its
unperturbed value of $-q$.  There is thus no way to bring it to the
value of $-c/D$, required by matching to the solution in the non-reacting
region.  For $c < 2\sqrt{D}$, however, $\phi$ passes
through zero, and so $\phi'/\phi$ diverges.  Thus, $\phi'/\phi$ crosses
$-c/D$ at some point.  The closer we are to $c=2\sqrt{D}$, the
more this crossing is pushed out to larger $y$, and the smaller the
value of $\phi$ is at the crossing. But $\phi$ at the crossing is just 
$\epsilon$, so we see why $c$ approaches $2\sqrt{D}$ from below
when $\epsilon$ is taken to zero.  We can understand in a simple way the
origin of the large, $O(\ln \epsilon)^{-2}$ correction to the
velocity.  Due to the degeneracy of the decay rate of the two solutions
of the linear operator for $c=2\sqrt{D}$, the steady-state solution
behaves for large $y$ as $\phi \sim Ay\exp(-y/\sqrt{D})$. If we then
examine the perturbative effect of a change in $c$ on this
solution, the front region is described by
\begin{equation}
\phi\sim Aye^{-y/\sqrt{D}}(1 +  \delta c \frac{y^2}{6D^{3/2}})
\end{equation}
which immediately implies that $y_c \sim \sqrt{D}\ln(1/\epsilon)$ and 
$\delta c \sim \sqrt{D}(\ln\epsilon)^{-2}$.  
Thus $\delta c$ is relatively large
because the first correction is not exponentially larger than
the steady-state term, as in GL, but only power-law larger.  Thus,
just as
the {\em stability} of the pure dying exponential behavior for $c>2\sqrt{D}$
gives rise to the structural {\em instability} of these solutions to
the introduction of the cutoff, so the very singular 
behavior of the velocity in the $\epsilon \to 0$ limit arises from the
relatively weak instability of the front behavior of the $c=2\sqrt{D}$
solution.

There is an intriguing correlation between the two properties investigated
by BD.  In essence, we can argue that the initial conditions,
which are responsible for the velocity selection in the Fisher equation,
act as a sort of effective cutoff beyond which the steady-state equation
no longer holds.  This cutoff is however time-dependent, and goes to
zero as more and more of the steady-state front is established.  Roughly
speaking, since the front is close to the steady-state front out to 
a distance $y_c$ of order $t^{1/2}$,
the effective cutoff, $\epsilon_{\text{eff}}$, is of order $\exp(-qy_c)$.  
Then, the velocity
is expected to be less than the steady-state velocity by an amount of
order $(\ln \epsilon_{\text{eff}})^{-2} \sim y_c^{-2} \sim 1/t$, which
is indeed what BD find.  The argument is not precise, as
indicated by the fact that the numerical prefactor in this relation
does not come out correctly.  Nevertheless, it does seem to capture
the correct physics.  It is amusing to note that a similar argument
can be made for GL.  Here the exponential approach with time of the velocity
to its steady-state value is correctly reproduced, but the exponent comes
out wrong by a factor of 4.  Thus, in both cases, it is as if the true 
effective cutoff is some power of our naive $\epsilon_{\text{eff}}$
estimated above.

We also note that the introduction of the cutoff has another effect
on the dynamics.  Instead of the $1/t$ convergence of the velocity
in the absence of the cutoff, the velocity now converges exponentially
to its long-time value.  For small cutoff, the velocity follows the
0-cutoff $1/t$ behavior up to some time, slowly increasing, and then 
sharply ``breaks'' when it nears the velocity of the cutoff model.
In terms of the picture presented above, the dynamics is insensitive
to the cutoff as long as the ``dynamic'' cutoff, $\epsilon_{\text{eff}}$,
is larger than $\epsilon$.  Eventually, $\epsilon_{\text{eff}}$ falls
below $\epsilon$, and the velocity exponentially approaches its asymptotic
value.  This behavior is exhibited in Fig. 4.  Also Of interest is the
time constant that governs the post-break time decay. The equation for the
stability operator, transformed to a Schroedinger-type representation
by eliminating the first-derivative term via a similarity transformation,
reads
\begin{equation}
-\Omega \delta \phi = -(\delta \phi)'' + \left[\frac{c^2}{4D} - 
\theta(y_c-y)(1 - 2\phi_0) - \delta(\phi_0(y)-\epsilon)(\phi_0)(1-\phi_0)\right]\delta\phi
\end{equation}
where $y_c$ marks the cutoff point, $\phi_0(y_c)=\epsilon$ and $\Omega$
is the decay rate of the perturbation.
For large $y$, small $\epsilon$, this becomes
\begin{equation}
-\Omega \delta \phi = -D(\delta \phi)'' + \left[\frac{c^2}{4D} -
\theta(y_c-y)  - \frac{D}{c}\delta(y-y_c)\right]\delta\phi
\end{equation}
We see that the flat region of the ``potential'' term at large $y$ with
value $c^2/4D - 1$ which was responsible for the onset of the
continuum at $\Omega=0$ is now transformed by the addition of a step
of height 1 at $y=y_c$, in addition to a $\delta$-function potential
at this point.  Thus, the continuum of modes is rendered discrete with
a spacing inversely proportional to the square of the width of the
potential.  This width is just $y_c$ for small $\epsilon$.  Using
the result from BD that $y_c=\pi/\sqrt{D}\ln(1/\epsilon)$, we find a
spacing proportional to $(\ln \epsilon)^{-2}$.  Thus, the time constant 
$1/\Omega$ diverges with $\epsilon$, but very slowly.

\begin{figure}
\centerline{\epsfxsize=3.25in \epsffile{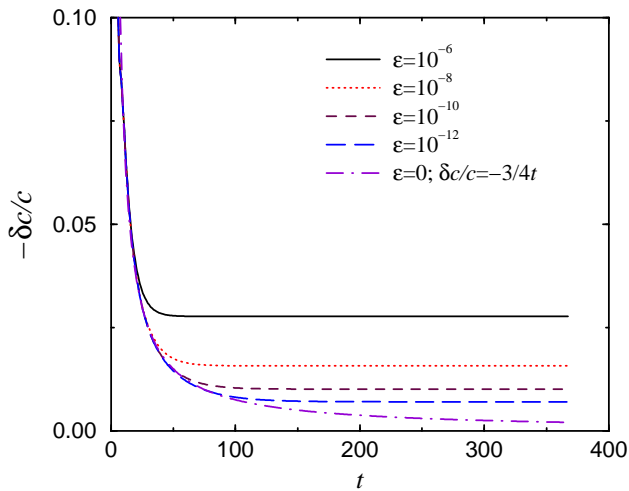}}
\caption{$-\delta c(t)/c$ vs. $t$ where $\delta c(t)\equiv c(t)-c$,
$c=c(t=\infty)=2\protect{\sqrt{D}}$ for the cutoff Fisher eqn., 
with $D=1/2$, $\epsilon = 10^{-6}$, $10^{-8}$, $10^{-10}$, $10^{-12}$, and
$0$.
The data for $\epsilon=0$ is indistinguishable from the BD prediction of
$\delta c/c=3/4t$.}
\end{figure}

\section{Stochastic Fisher model}
BD offer the conjecture that in a stochastic model of US type,
the discreteness of the
particles gives rise to an effective cutoff.  The point here is since the $\epsilon$-induced correction
to the velocity is so large, it should overwhelm any other effect.  
It should be noted, however, that the
point is not completely obvious.  It is true that each
realization has some point beyond which no interactions occur.
However, if one considers the ensemble average, there is no
such sharp cutoff.  Since the average velocity is a function of
the ensemble, one might have cause to question BD's conjecture.

BD found support for their conjecture from simulations
of a model related to directed polymers.  
Due to the importance of this point, we have sought further
evidence via a simulation whose
mean-field limit is precisely the Fisher equation.  The model is a
slight modification of an infection model studied by Blumen, et. al.
\cite{Blumen}.  In our model, we introduce an infectious (black) particle
into a sea of white particles.  All particles move randomly, with
no constraint on the number of particles at any one site.  When a black
particle and a white particle are on the same site in a given time step, 
$\Delta t$,
they have a small probability,  $k\Delta t$,  
of interacting during this time. If they
do interact, the white particle is ``infected'' and turns black, and may
go on to infect additional particles at later times.  
If the black particle is introduced at one end of the system, 
then a front is established which
invades the white phase at a velocity which on average is constant in time.
It is particularly convenient to work with a parallel version of this
algorithm, using a binomial random number generator to determine the
number of interactions and hops at each site during each time step.  This
way we can take the average number of particles per site, $N$, to be huge, with
essentially no cost in performance.  This form is also nice, since if
the binomial random number generator is replaced by the mean of the
binomial distribution, we explicitly obtain a discretized form of the
Fisher equation.  

We have performed extensive simulations of this model, measuring the
average velocity of the front.  The results are presented in Fig. 5.
We see that they are consistent with a $(\ln N)^{-2}$ behavior.  The
prefactor, however, differs significantly from that of the mean-field
version, supplemented by a BD-type cutoff of $\epsilon=1/N$.  (The 
much smaller 10\% discrepancy
between the mean-field prefactor and the BD prediction of $\pi^2/2$ is
the result of the finite lattice and time step $\Delta t$.)
This reiterates the findings of BD for their discrete model where
the prefactor also did not match the prediction.  Since any simple rescaling
of the cutoff would not affect the prefactor, this effect is not trivial.
Evidently the effect of the discreteness is more subtle, and
similar to the effective cutoff imposed by the initial condition,
involves an effective cutoff that is a power of the naive one.
This point merits further investigation.

\begin{figure}
\centerline{\epsfxsize=3.25in \epsffile{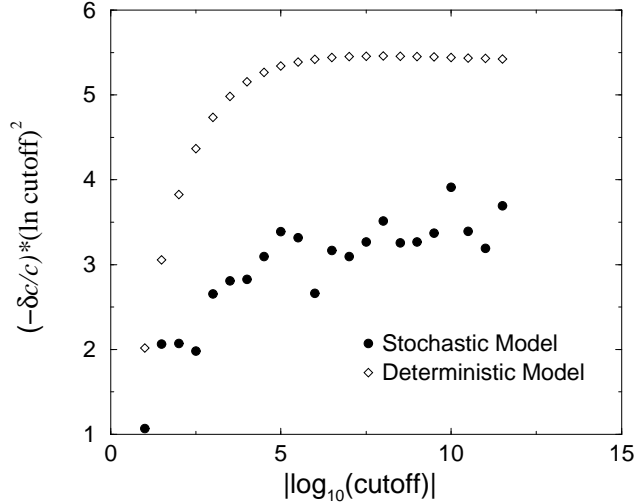}}
\caption{$(-\delta c/c)(\ln\text{cutoff})^2$ vs. $|\log_10(\text{cutoff})|$
for the stochastic Fisher model and its cutoff deterministic 
(mean-field) counterpart. For the stochastic Fisher model, the cutoff is $N$, the average number of 
particles per site.  For the deterministic counterpart, the cutoff is
$\epsilon=1/N$. In both models, the parameters are $k=0.1/N$, $\Delta t=0.1$,
$D=0.5$, and the lattice spacing, $dx=1$.}
\end{figure}
\begin{figure}
\centerline{\epsfxsize=3.25in \epsffile{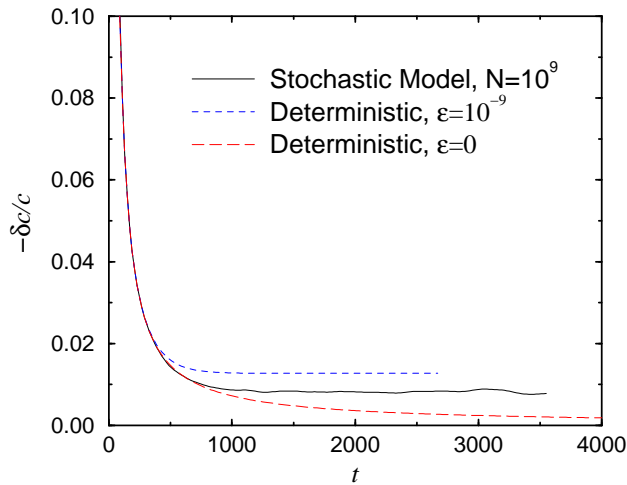}}
\caption{$(-\delta c(t)/c)$ vs. $t$
for the stochastic Fisher model and its deterministic 
(mean-field) counterpart, with and without a cutoff, where $\delta c(t)$
is as in Fig. 4. For the stochastic 
Fisher model, $N=10^9$.  For the cutoff deterministic counterpart, 
the cutoff is
$\epsilon=1/N$. The other parameters are as in Fig. 6}
\end{figure}

Not only the $N$-dependence of the velocity, but also its time-dependence
mirrors that of the cutoff Fisher equation.  As seen in Fig. 6, the 
velocity initially falls
as in the $\epsilon=0$ deterministic model, as $t^{-1}$ but eventually
breaks and converges exponentially to its asymptotic value.
The break point is past that of the model with the naive cutoff, 
$\epsilon=1/N$.  This is consistent with what we found above, that the
asymptotic velocity is {\em closer} to the Fisher result than one would
naively expect.

The question remains why BD's conjecture works, given the reservations 
expressed above.  It is tempting to
suggest that the answer lies in a correct definition of the ensemble
average.  If the ensemble average is taken in the lab frame, it is
clear that the ensemble-averaged reaction density has no cutoff.  Even
if the average is taken relative to the bulk front (defining the
front position where the concentration crosses $N/2$, say), reactions
still have a non-zero probability of occuring arbitrarily far in front
of the front.  However, if the average is
taken with respect to the frame defined by the lead particle position, 
then indeed the averaged concentration so computed
has a sharp cutoff, and no reactions occur beyond this point.  
Such a definition was found to be crucial in studying
the ensemble average of a model of (transparent) diffusion-limited-aggregation 
in one dimension \cite{DLA}.  It would appear that here too a sensible
description of the ensemble average requires this kind of averaging
procedure.

\section{Propagation into an Unstable State: Ginzburg-Landau}
Given that the GL equation has an unstable state, we can also consider
propagation into the unstable state.  For ease of comparison to 
the Fisher equation, we will set the unstable state to $\phi=0$ and
the stable states to $\phi=1$ and $\phi=-1/\alpha$, so that the GL
equation reads
\begin{equation}
\label{glII}
\dot\phi = D \phi'' + \phi(1-\phi)(1 + \alpha\phi)
\end{equation}
Clearly, for $\alpha=0$, this simply reduces to the Fisher eqn. with its
asymptotic velocity of $c=2\sqrt{D}$.  In fact, for all $\alpha$, the
equation possesses steady-state propagating solutions interpolating between
$\phi=1$ and $\phi=0$ for all velocities, just
as in the Fisher equation, since the $\phi=0$ state is unstable.
Furthermore, since the linearization around $\phi=0$ is independent of
$\alpha$, one would naively expect that the initial condition would again
select the degenerate $c=2\sqrt{D}$ solution for all $\alpha$.  
However, Ben-Jacob, et. 
al. (BJ) \cite{Ben-Jacob} 
note that this expectation is only met for $\alpha\le 2$.  For $\alpha>2$,
on the other hand, the selected velocity is greater than this.  As BJ
note, this is connected to the existence of a special velocity 
$\hat c(\alpha)=(\alpha+2)\sqrt{D/2\alpha}$
for $\alpha>2$.  They show that, for this velocity, the solution does not 
have the spatial decay rate $q=(c-\sqrt{c^2 - 4D})/2D$ typical of all other
velocities.  Rather, the spatial decay rate is faster, that of the typically
sub-dominant solution of the linearized steady-state equation: 
$\hat q = (c+\sqrt{c^2 - 4D})/2D$.  Of course, this behavior is unstable,
and so is associated with a unique (or, in general, a discrete set of) 
velocity. This situation, which BJ have entitled as ``nonlinear'' or 
``Type II'' marginal
stability, and we will label as US-II, is in a sense intermediate
between the MS case discussed in Section 2, and
the standard US case as exemplified by the Fisher eqn. in Section 3.   

BJ note that for $\alpha>2$, $\hat c(\alpha)$ is the
selected velocity.  We can understand this in light of our previous
discussions.  Since the decay rate for this solution is $\hat q$, not $q$,
the velocity of the precursor is positive (in the co-moving frame), with
value $c^*=\sqrt{c^2 - 4D}=(\alpha-2)\sqrt{D/2\alpha}$.  
For all other solutions with $c>2\sqrt{D}$, the precursor has negative 
velocity, and
so these solutions cannot be constructed dynamically.  The $c=2\sqrt{D}$
solution is ruled out, since, as pointed out by Ben-Jacob, et. al., this
solution is unstable (as are all solutions with $c<\hat c(\alpha)$).  Thus
like the MS case, the precursor has positive velocity relative to the front.
However, unlike in the MS case, where this relative velocity is always 
larger than
the front velocity $c$, here this relative velocity can be arbitrarily small.
Thus, as in the MS case, the approach of the front velocity to
its asymptotic value is exponential, but the time constant, being inversely 
proportional to $(c^*)^2$, becomes arbitrarily long as $\alpha$ approaches
2. 

We now turn to the effect of a cutoff on the selected velocity.  As in the
MS case, the instability of the large $y$ behavior of the $\hat c$ solution 
with respect to a change in $c$ indicates that a solution of the
cutoff problem can only be constructed close to $\hat c$.
Setting $c=\hat c +\delta c$, the dominant
behavior for $\phi$ is, to leading order,
\begin{equation}
\phi = A e^{- \hat q y} + B \delta c e^{- q y}
\end{equation}
Even though both modes are decaying exponentials, unlike the MS case (Eq. 
(\ref{mstail})), nevertheless it is true that as with MS, and unlike Fisher,
the correction term is exponentially larger than the original solution.
Working out the matching to the cutoff region, the correction to the
velocity is seen to scale as 
\begin{equation}
\delta c \sim \epsilon^{1 - q/\hat q}=\epsilon^{(\alpha-2)/\alpha}
\end{equation}
This behavior is exemplified in Fig. 7.
Again, as in MS, the correction is power-law in $\epsilon$, but now
the power is always less than 1, and goes to 0 as $\alpha$ approaches 2.
At the other extreme, the power approaches 1 as $alpha \to \infty$, whereupon
the metastable state at $\phi=-1/\alpha$ merges with the unstable
state at $\phi=0$, rendering it marginally (meta-)stable. This then
naturally matches onto the limit of the MS case studied in Sec. II, when
$a \to 0$.  In general, then,, the cutoff dependence for US-II is more 
singular than MS but less singular than
US.  

\begin{figure}
\centerline{\epsfxsize=3.25in \epsffile{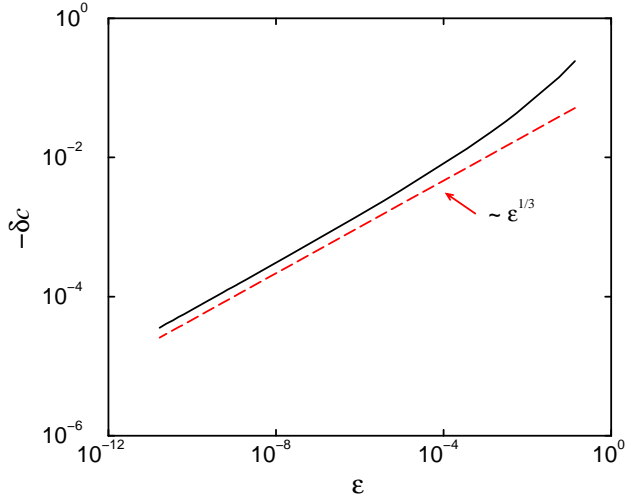}}
\caption{$-\delta c$ vs. $\epsilon$, where $\delta c=c(\epsilon)-
c(\epsilon=0)$
for US propagation in the GL equation, Eq. (\ref{glII}), with $D=1/2$, 
$\alpha=3$.  The
theoretical scaling prediction $\delta c \propto \epsilon^\beta$,
$\beta = (\alpha-2)/\alpha=1/3$ is also shown.}
\end{figure}

Lastly, let us discuss the nature of the stability spectrum.  Ben-Jacob, et.
al. in their discussions of stability do not give
the boundary-conditions necessary to defined the spectrum of the stability
operator.  It seems clear to us, however, that the appropriate definition
in the cases we have examined above seems clear.  If we transform the
problem into Schroedinger form, as discussed above, we know what boundary
conditions give rise to a complete set of states.  This completeness
is crucial if we wish to decompose some initial condition via this
set of states and thus integrate the linear problem forward in time.
In particular, we must demand that any discrete state vanish exponentially
at both $\pm \infty$.  This leads to the striking conclusion that the
stability spectrum of the Fisher equation about any $c>2\sqrt{D}$ solution
does not include the translation zero mode!  This mode, after the similarity
transformation, diverges at $\infty$.  This fact in and of itself would
lead one to suspect that these solutions are illegitimate.  Only
at $c=2\sqrt{D}$ does the operator recover, at least in some weak
sense, a zero mode, since in this case the continuum starts at 0.

The
case of propagation into the unstable phase in the GL equation is 
very illuminating in this context.  For $\alpha<1$, 
numerical work indicates that the spectrum consists solely of the
continuum modes, extending upward from $\Omega=c^2/4D - 1$, with no
discrete modes. Of course, this has to be, or the only realizable 
solution for this range of $\alpha$, the $c=2\sqrt{D}$ one, would be unstable.
However, for $\alpha>2$, a discrete mode emerges
out of the continuum.  This mode has
a positive $\Omega$ for large $c$, and crosses 0 exactly at $\hat c$.  This
is guaranteed since the translation mode for the $\hat c$ solution is
a legitimate eigenmode of the Schroedinger equation, due to the fast spatial
falloff of this solution. For $c<\hat c$, the mode has negative
$\Omega$ and so is unstable.  Again, the translation mode is illegitimate,
and this is the only discrete mode.  Thus, the selected solution is the
only one which has a legitimate translation mode in its stability spectrum.
This fact is related to the structural stability of the $\hat c$ solution 
in the
presence of a cutoff.  When discussing the spectrum of a Schroedinger
operator in an infinite space, it is traditional to put the system in
a finite box.  This box is another kind of cutoff, and only the $\hat c$
solution survives for a large but finite box.  This may be a more appropriate
way of understanding the ``marginal stability'' of the selected solution,
namely the requirement that the stability operator of a physically
realizable solution possess a translation zero mode.

\section{Conclusions}

In this paper, we have seen how analyzing the time-development of the
front in the MS case sheds a new light on many aspects of front propagation.
In particular, we have shown how the MS front develops through the propagation
of a diffusively spreading precursor front which moves at least 
twice as fast as the front itself.  Ahead of this precursor, the field
falls off in the  Gaussian manner typical of a pure diffusion problem.
Behind, the exponential tail of a propagating front is established.  This
precursor is a essential element of propagating fronts, and its nonexistence
in the US case explains the dynamical velocity selection seen therein.

We have also seen how the precursor follows immediately from the structure 
of the linear operator ahead of the
front.  This structure is intimately connected also
with the structural stability of the problem to the introduction of a cutoff,
as well as to the nature of the stability spectrum.  This cutoff can
arise either from the finiteness of the system, the discreteness of the
underlying reacting system, or even from the initial conditions. 
The two issues of cutoffs and the stability spectrum are themselves
interdependent, as we have seen, and in turn tie back to and 
give a deeper understanding
to the original picture of marginal stability.  

\acknowledgements
DAK thanks Herbert Levine for useful conversations.
DAK acknowledges the support of the Israel Science Foundation.  LMS 
acknowledges the support of the NSF, Grant No. DMR 94-20335, and the
hospitality of the Weizmann Institute, where this work was begun.

\end{document}